\documentclass[useAMS,usenatbib]{mn2e}
\usepackage{amssymb}
\usepackage{amsmath}
\usepackage{graphicx}

\title[]{Application of a screened Coulomb potential in fitting the DAV star HS 0507+0434B}
\author[Y. H. Chen]{Y. H. Chen$^{1,2,3}$\thanks{E-mail: yanhuichen1987@126.com}\\
$^{1}$Institute of Astrophysics, Chuxiong Normal University, Chuxiong 675000, China\\
$^{2}$School of Physics and Electronical Science, Chuxiong Normal University, Chuxiong 675000,China\\
$^{3}$Key Laboratory for the Structure and Evolution of Celestial Objects, Chinese Academy of Sciences, P.O. Box 110, Kunming 650011, China\\}

\begin{document}

\date{Accepted: }

\pagerange{\pageref{firstpage}--\pageref{lastpage}} \pubyear{????}

\maketitle

\label{firstpage}

\begin{abstract}

We use \texttt{WDEC} to evolve grids of DAV star models adopting the element diffusion scheme with pure and screened Coulomb potential effect. The core compositions are thermonuclear burning results which are derived from \texttt{MESA}. \texttt{MESA} yields composition profiles that the version of WDEC used in this work could not accommodate (most notably the presence of helium in the core of the model). According to the theory of rotational splitting, Fu et al. identified 6 triplets for a DAV star HS 0507+0434B based on 206\,h photometric data. The grids of DAV star models are used to fit the 6 reliable $m$ = 0 modes. When adopting the screened Coulomb potential, we obtain a best fitting model of log($M_{\rm He}/M_{\rm *}$) = -3.0, log($M_{\rm H}/M_{\rm *}$) = -6.1, $T_{\rm eff}$ = 11790\,K, $M_{\rm *}$ = 0.625\,$M_{\odot}$, log$g$ = 8.066, and $\sigma_{RMS}$ = 2.08\,s. Compared with adopting the pure Coulomb potential, the value of $\sigma_{RMS}$ is improved by 34\%. The study may provide a new method for the research of mode trapping properties.

\end{abstract}


\section{Introduction}

White dwarfs are the last observable evolutionary stage of almost 98\% of all stars (Winget \& Kepler 2008). It is of wide and universal significance to study white dwarfs. White dwarfs have the mass of the order of the sun and the volume of the order of the earth, so they have large density and super gravity. White dwarfs are natural laboratories for the extreme physical laws of super gravity and degenerate electron. The electron degenerate core and the ideal gas atmosphere are the basic structures of white dwarfs. The white dwarfs with hydrogen (H) rich atmosphere are called DA type white dwarfs. DB type white dwarfs have rich helium (He) atmosphere and DO type white dwarfs have rich ionized-He atmosphere. Carbon (C) and oxygen (O) lines are also present in the atmosphere for DO type white dwarfs, such as PG 1159-035 (Werner, Heber \& Hunger 1991). The process of thermonuclear burning basically stops, and the cooling (together with contraction early on) dominates the evolution of white dwarfs. Along the white dwarf cooling branch, there are DOV (DO type variable white dwarfs), DBV and DAV instability strips. The instability strip ranges from $\sim$170,000\,K to $\sim$75,000\,K for DOV stars, $\sim$29,000\,K to $\sim$22,000\,K for DBV stars, $\sim$12,270\,K to $\sim$10,850\,K for DAV stars (Winget \& Kepler 2008).

Asteroseismology is a more progressive and powerful tool to probe the inner structure of stars. It is very important to have reliable mode identifications and physical theoretical models. In recent years, based on large telescopes, WET (Whole Earth Telescope) runs and multi-site observations, and space missions such as Kepler and TESS, asteroseismology on white dwarfs has made a considerable progress.

There are many progresses in the study of physical theoretical models. The White Dwarf Evolution Code (\texttt{WDEC}) calculates the white dwarf cooling processes (Bischoff-Kim \& Montgomery 2018). As an open source, \texttt{WDEC} is very convenient to calculate grids of white dwarf models with artificial core compositions. \texttt{LPCODE} is used to make full evolutionary white dwarf models from the main-sequence stage. During the white dwarf evolution, the time-dependent element diffusion effect is added (see Althaus et al. 2010, Romero et al. 2012 and their recent papers). The Modules for Experiments in Stellar Astrophysics (\texttt{MESA}) can evolve a star from the pre-main sequence stage to the white dwarf stage (Paxton et al. 2011). It is an open source software. Chen \& Li (2014a) added the core compositions of white dwarfs evolved by \texttt{MESA} to \texttt{WDEC} to evolve grids of white dwarf models. The thermonuclear burning core is more physical than previous artificial core in \texttt{WDEC}. Su et al. (2014) incorporated the scheme of Thoul, Bahcall \& Loeb (1994) into \texttt{WDEC} to treat the element diffusion of H, He, C, and O. The equation of Coulomb logarithm (Eq.\,(9) in Thoul, Bahcall \& Loeb 1994) is for a pure Coulomb potential with a cutoff at the Debye radius. White dwarfs are extremely compact objects with super gravity fields. In the high-density stellar plasma environment, the screened Coulomb potential may be more suitable. Paquette et al. (1986) provided a better description of the white dwarf plasma with a screened Coulomb potential.

Considering the physical effect of the screened Coulomb potential, it is possible for us to move towards the direction of precision asteroseismology. Chen (2018) first applied the screened Coulomb potential to fit a DBV star PG 0112+104. The composition transition zone is closely related to the screened Coulomb potential. However, the effect of screened Coulomb potential on DAV stars has not been studied.

Based on multi-site observation campaigns in 2007 and from 2009 December to 2010 January, Fu et al. (2013) obtained 206 h of photometric time series data for DAV star HS 0507+0434B. Studying the stellar rotation period ($P_{\rm rot}$) and the frequency splitting value ($\delta\nu_{k,l}$), Brickhill (1975) derived an approximate formula of
\begin{equation}
m\delta\nu_{k,l}=\nu_{k,l,m}-\nu_{k,l,0}=\frac{m}{P_{\rm rot}}(1-\frac{1}{l(l+1)}).
\end{equation}
\noindent In Eq.\,(1), $k$ is the radial overtone, $l$ is the spherical harmonic degree, and $m$ is the azimuthal number. The eigen-frequencies will be split into 2$l$+1 due to the stellar rotation effect. For $l$ = 1 modes, the frequencies show triplets. Based on the relationship, Fu et al. (2013) identified 6 triplets for HS 0507+0434B with an average $\delta\nu_{k,1}$ of 3.59$\pm$0.57$\mu$Hz. The average amplitude ratio between the $m$=$\pm$1 components and the $m$=0 components is 1.98 for 4 triplets. Fu et al. (2013) reported that the angle of the rotation axis to the line of sight was close to 70$^{\circ}$ because of the theoretical study of Pesnell (1985). In addition to the 6 triplets, Fu et al. (2013) also identified one independent mode of 999.7\,s. They also identified 10 further signals with low amplitude which were more uncertain detections at the limit of selection criterion and/or which were close to possible liner combinations but did not fulfil the 3$\times$$\sigma$ criterion. The 6 $m$ = 0 components are very reliable. We use the 6 $m$ = 0 components to constrain the fitting models adopting pure and screened Coulomb potential. Based on the best-fitting model, the fittings of the independent mode of 999.7\,s and some further signals and linear combinations will be discussed. In Sect. 2, we show the input physics and model calculations. In Sect. 3, we show the asteroseismological study on HS 0507+0434B. We discuss the selection of the best fitting model on HS 0507+0434B in Subsect. 3.1. In Subsect. 3.2, we discuss the effect of screened Coulomb potential. The comparisons between the best fitting model and the previous work are displayed in Subsect. 3.3. At last, we give a discussion and conclusions in Sect. 4.

\section{Input Physics and Model Calculations}

In this work, we used \texttt{MESA} version 4298 to evolve white dwarf models in order to obtain base chemical profiles. We incorporated those profiles into WDEC to perform asteroseismic fits. For the He/H and C/He transition zones, we did not take the diffusion equilibrium profiles in \texttt{WDEC}. Our pure Coulomb potential DAV star models are calculated by \texttt{WDEC} taking the element diffusion scheme of Thoul, Bahcall \& Loeb (1994) which was added into \texttt{WDEC} by Su et al. (2014). Considering the stability of numerical calculation and the selection of internal boundary conditions, the element diffusion scheme calculates the C/He transition zones and more external profiles. The element diffusion scheme does not calculate the chemical compositions at the central core. The element diffusion effect mainly affects the profiles of He/H and C/He transition zones.

The scheme of Thoul, Bahcall \& Loeb (1994) was used to study the element diffusion effect in the solar interior. The equation of Coulomb logarithm (Eq.\,(9) in Thoul, Bahcall \& Loeb 1994) is for a pure Coulomb potential with a cutoff at the Debye radius. For white dwarfs, we replace the pure Coulomb potential into a Debye-like potential of
\begin{equation}
V_{st}(x)=\frac{Z_{s}Z_{t}e^{2}}{r}e^{-(r/\lambda)}.
\end{equation}
\noindent In Eq.\,(2), $Z_{s}$, $Z_{t}$ are particle charges, $r$ is the distance, and $\lambda$ = max ($\lambda_{D}$, $a_{0}$). The parameter $\lambda_{D}$ is the Debye length and $a_{0}$ is the mean ionic distance. According to the research of Muchmore (1984) and Cox, Guzik \& Kidman (1989), the Burgers equations for momentum and energy conservation (Eq.\,(12) and (13) in Thoul, Bahcall \& Loeb (1994)) are revised. With equations (1, 2, 22-25) in Muchmore (1984), we revised the equations (9, 12, 13) in Thoul, Bahcall \& Loeb (1994). The DAV star models evolved by the above method are considered as taking the element diffusion effect with the screened Coulomb potential into account. For more details, see Chen (2018). We add the thermonuclear burning core compositions to \texttt{WDEC} and we use \texttt{WDEC} to evolve grids of DAV star models. The core compositions are column data of mass, radius, luminosity, pressure, temperature, entropy, and C abundances. The maximum C abundance (about 0.65) is truncated as the boundary of the core compositions. The oxygen abundance equals 1.0 minus the carbon abundance in the core. The seed models are exactly same with Table 4 in Chen \& Li (2014a). Then, the DAV star models have core compositions of quasi-nuclear-burning results. The \texttt{WDEC} can not calculate EOS of three elements. In order to avoid an interface of He/C/O, we set the carbon abundance to 1.0 at the surface of the core. This results in a C/He interface.

\texttt{WDEC} computes white dwarf models. DAV star models are evolved by an older version of \texttt{WDEC}. \texttt{WDEC} was first written by Schwarzschild and subsequently modified by many authors, such as Kutter \& Savedoff (1969), Lamb \& van Horn (1975), and Wood (1990). \texttt{WDEC} is a fast and versatile code designed for the white dwarf evolutions. There is not nuclear reaction, mass-loss, and previous stellar evolutions for \texttt{WDEC}. Lamb (1974) derived the equation of state (EOS) in the degenerate core and Saumon et al. (1995) derived the EOS in the radiant shell. The opacities are from Itoh et al. (1983, 1984). For more details, see Bischoff-Kim \& Montgomery (2018). We use the code to evolve grids of DAV star models. The core composition profiles are derived from \texttt{MESA}, but have to be modified for use a version of \texttt{WDEC} that does not accommodate He in the core. For the He/H and C/He transition zones, we take the element diffusion scheme with pure and screened Coulomb potential into account. Bergeron et al. (1995) reported that a model with $\alpha$ = 0.6 provided an excellent internal consistency between ultraviolet and optical temperatures. We set the mixing length parameter $\alpha$ to 0.6.

The grid parameters, initial-version of grid size and grid steps are shown in the first three columns in Table 1. The middle-steps and refined steps will be discussed later according to the initial-best fitting models. With the pulsation code of Li (1992a, b), we numerically solve the full equations of linear and adiabatic oscillation on these DAV star models. The calculated modes are used to fit the observed modes.

\begin{table}
\begin{center}
\begin{tabular}{lllll}
\hline
Parameters                  &initial               &initial     &middle        &refined                             \\
                            &-size                 &-steps      &-steps        &-steps                              \\
\hline
$M_{*}$/$M_{\odot}$         &0.56to0.72            &0.01        &0.005         &0.005                               \\
$T_{\rm eff}$(K)            &10800to12600          &200         &50            &10                                  \\
log($M_{\rm He}/M_{\rm *}$) &-2.0to-4.0            &0.5         &0.5           &0.1                                 \\
log($M_{\rm H}/M_{\rm *}$)  &-4.0to-10.0           &1.0         &0.5           &0.1                                 \\
\hline
\end{tabular}
\end{center}
\caption{The grid size and steps table.}
\end{table}

A root-mean-square residual ($\sigma_{RMS}$) is calculated by
\begin{equation}
\sigma_{RMS}=\sqrt{\frac{1}{n} \sum_{n}(P_{\rm obs}-P_{\rm cal})^2}.
\end{equation}
\noindent In Eq.\,(3), $n$ is the number of observed modes (6 for HS 0507+0434B). The smaller the value of $\sigma_{RMS}$, the better the fitting results. The calculated modes are used to fit the 6 $m$ = 0 modes. Some initial-best fitting models will be selected. Then, the steps of grid parameters will be reduced around these initial-best fitting models. The theoretical modes are calculated and the Eq.\,(3) is used again. At last, a best fitting model will be selected.

\section{Asteroseismic study}

With the grid of DAV star models, we do the asteroseismological study of HS 0507+0434B. We try to use the asteroseismological study to evaluate the application of screened Coulomb potential in DAV star HS 0507+0434B. A reliable mode identification is essential for the study. We only use the 6 $m$ = 0 modes of HS 0507+0434B from Fu et al. (2013) to constrain the fitting models.

 \subsection{Selection of the best fitting model}

\begin{figure}
\begin{center}
\includegraphics[width=9.0cm,angle=0]{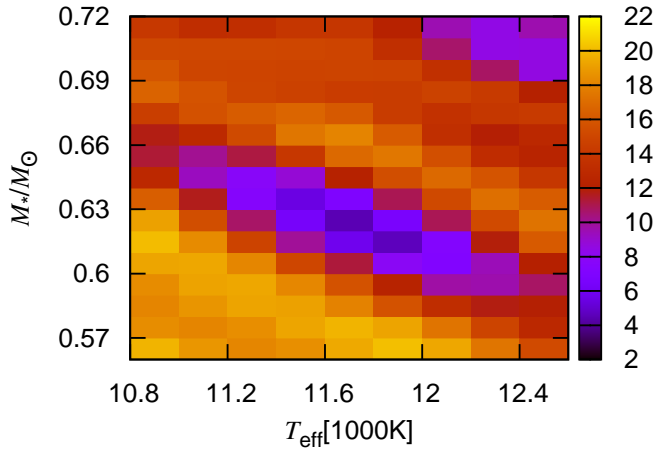}
\end{center}
\caption{The color residual diagram for fitting HS 0507+0434B with DAV star models adopting the screened Coulomb potential. For the models, log($M_{\rm He}/M_{\rm *}$) is -3.0 and log($M_{\rm H}/M_{\rm *}$) is -6.0. The colors indicate the values of $\sigma_{RMS}$.}
\end{figure}

\begin{table*}
\begin{center}
\begin{tabular}{llllllllllllllll}
\hline
ID          &$T_{\rm eff}$  &$M_{*}$                &log($M_{\rm H}/M_{*}$)  &log($M_{\rm He}/M_{*}$)&$\sigma_{RMS}$\\
            &(K)            &($M_{\odot}$)          &                        &                       &(s)           \\
\hline
$1(p)_{ini}$&11600          &0.63                   &-6.0                    &-3.0                   &3.48          \\
$1(p)_{mid}$&11700          &0.625                  &-6.0                    &-3.0                   &3.21          \\
$1(p)_{ref}$&11800          &0.625                  &-6.1                    &-3.0                   &3.09          \\
$1(s)_{ini}$&11800          &0.62                   &-6.0                    &-3.0                   &2.98          \\
$1(s)_{mid}$&11800          &0.62                   &-6.0                    &-3.0                   &2.98          \\
$1(s)_{ref}$&11790          &0.625                  &-6.1                    &-3.0                   &2.08          \\
$2(p)_{ini}$&12000          &0.72                   &-9.0                    &-3.5                   &3.50          \\
$2(p)_{mid}$&12050          &0.72                   &-9.0                    &-3.5                   &3.20          \\
$2(p)_{ref}$&12060          &0.72                   &-9.0                    &-3.5                   &3.19          \\
$2(s)_{ini}$&12000          &0.72                   &-9.0                    &-3.5                   &3.33          \\
$2(s)_{mid}$&12050          &0.72                   &-9.0                    &-3.5                   &3.05          \\
$2(s)_{ref}$&12050          &0.72                   &-9.0                    &-3.5                   &3.05          \\
$3(p)_{ini}$&11200          &0.65                   &-6.0                    &-3.5                   &3.78          \\
$3(p)_{mid}$&11250          &0.65                   &-6.0                    &-3.5                   &3.45          \\
$3(p)_{ref}$&11180          &0.65                   &-5.9                    &-3.5                   &3.23          \\
$3(s)_{ini}$&11200          &0.65                   &-6.0                    &-3.5                   &3.30          \\
$3(s)_{mid}$&11200          &0.65                   &-6.0                    &-3.5                   &3.30          \\
$3(s)_{ref}$&11230          &0.65                   &-6.0                    &-3.4                   &3.25          \\
$4(p)_{ini}$&12000          &0.59                   &-6.0                    &-4.0                   &4.91          \\
$4(p)_{mid}$&12050          &0.595                  &-6.0                    &-4.0                   &4.20          \\
$4(p)_{ref}$&12030          &0.59                   &-6.0                    &-4.0                   &4.14          \\
$4(s)_{ini}$&12000          &0.59                   &-6.0                    &-4.0                   &3.83          \\
$4(s)_{mid}$&11950          &0.595                  &-6.0                    &-4.0                   &3.78          \\
$4(s)_{ref}$&12050          &0.595                  &-6.0                    &-3.9                   &3.71          \\
\hline
\end{tabular}
\end{center}
\caption{Table of best fitting models with initial-steps, middle-steps, and refined steps. There are four groups of initial-best fitting models. The model of ID number $1(s)_{ref}$ is the last best fitting model with $\sigma_{RMS}$ = 2.08\,s.}
\end{table*}

The grids of DAV star models are produced taking element diffusion effect with pure and screened Coulomb potential into account. The theoretical periods are calculated and used to fit the 6 $m$ = 0 modes. In Fig. 1, we show the color residual diagram for fitting HS 0507+0434B with DAV star models adopting the screened Coulomb potential. For the models, log($M_{\rm He}/M_{\rm *}$) is -3.0 and log($M_{\rm H}/M_{\rm *}$) is -6.0. The colors indicate the values of $\sigma_{RMS}$. In Fig. 1, we can see that an initial-best fitting model has $T_{\rm eff}$ = 11800\,K, and $M_{\rm *}$ = 0.62\,$M_{\odot}$. We obtain four groups of initial-best fitting models with $\sigma_{RMS}$ less than 5.0\,s marked as $1(p)_{ini}$/$1(s)_{ini}$ to $4(p)_{ini}$/$4(s)_{ini}$ in Table 2.

Around the initial-best fitting models, we reduce the grid steps to the middle-steps and then refined-steps in Table 1. Those models with smallest values of $\sigma_{RMS}$ are shown in Table 2. At last, we obtain a best fitting model with log($M_{\rm He}/M_{\rm *}$) = -3.0, log($M_{\rm H}/M_{\rm *}$) = -6.1, $T_{\rm eff}$ = 11790\,K, $M_{\rm *}$ = 0.625\,$M_{\odot}$, and $\sigma_{RMS}$ = 2.08\,s, marked as $1(s)_{ref}$ in Table 2. The value of $\sigma_{RMS}$ for the best fitting model is much smaller than that of other models.

In Table 2, the values of $\sigma_{RMS}$ are improved by reducing the grid steps. For the screened Coulomb potential evolving scenario, $\sigma_{RMS}$ is 2.98\,s for the model of $1(s)_{ini}$ and 2.08\,s for the model of $1(s)_{ref}$. Therefore, the refined grid steps are very important for model fittings.

In Table 3, we show the detailed fitting results with our best fitting model. The model parameters are log($M_{\rm He}/M_{\rm *}$) = -3.0, log($M_{\rm H}/M_{\rm *}$) = -6.1, $T_{\rm eff}$ = 11790\,K, $M_{\rm *}$ = 0.625\,$M_{\odot}$, and log$g$ = 8.066. For the pure Coulomb potential evolving scenario, $\sigma_{RMS}$ is 3.15\,s. For the screened Coulomb potential evolving scenario, $\sigma_{RMS}$ is 2.08\,s. The first column shows the 6 observed modes ($P_{obs}$). The second/4th column shows the calculated modes for the pure ($P_{cal(p)}$)/screened ($P_{cal(s)}$) Coulomb potential evolving scenario. The third/5th column shows the corresponding $P_{obs}$ minus $P_{cal(p)}$/$P_{cal(s)}$. For the pure Coulomb potential evolving scenario, the observed mode of 655.9\,s is fitted by a mode of 662.28\,s with an error of $\,$-6.38\,s. However, it is fitted by a mode of 658.71\,s with an error of $\,$-2.81\,s when adopting the screened Coulomb potential evolving scenario. For the mode of 655.9\,s the fitting error is from $\,$-6.38\,s to $\,$-2.81\,s, 56\% improved. Fitting the 6 observed modes, $\sigma_{RMS}$ is from 3.15\,s to 2.08\,s, 34\% improved.

\begin{table}
\begin{center}
\begin{tabular}{lllll}
\hline
$P_{obs}$    &$P_{cal(p)}$          &$P_{obs}$-$P_{cal(p)}$        &$P_{cal(s)}$ &$P_{obs}$-$P_{cal(s)}$              \\
\hline
(s)          &(s)                   &(s)                           &(s)          &(s)                                 \\
\hline
355.3        &355.62                &$\,$-0.32                     &354.47       &$\,$ 0.83                           \\
445.3        &445.04                &$\,$ 0.26                     &445.22       &$\,$ 0.08                           \\
556.5        &552.57                &$\,$ 3.93                     &553.05       &$\,$ 3.45                           \\
655.9        &662.28                &$\,$-6.38                     &658.71       &$\,$-2.81                           \\
697.6        &699.29                &$\,$-1.69                     &699.42       &$\,$-1.82                           \\
748.6        &749.34                &$\,$-0.74                     &747.15       &$\,$ 1.45                           \\
\hline
$\sigma_{RMS}$&3.15\,s              &                              &2.08\,s      &                                    \\
\hline
\end{tabular}
\end{center}
\caption{The detailed fitting results with pure and screened Coulomb potential models. The model parameters are log($M_{\rm He}/M_{\rm *}$) = -3.0, log($M_{\rm H}/M_{\rm *}$) = -6.1, $T_{\rm eff}$ = 11790\,K, $M_{\rm *}$ = 0.625\,$M_{\odot}$, and log$g$ = 8.066. $P_{obs}$ is the observed modes, $P_{cal(p)}$ is the calculated modes with the pure Coulomb potential, and $P_{cal(s)}$ is the calculated modes with the screened Coulomb potential.}
\end{table}

With the screened best-fitting model $1(s)_{ref}$ in Table 2, we show the detailed fitting results for the independent mode of 999.7\,s, some further signals, and some linear combinations identified by Fu et al. (2013) in Table 4. We take $\delta\nu_{k,1}$=3.59$\mu$Hz and $\delta\nu_{k,2}$=5.98$\mu$Hz ($\delta\nu_{k,1}$ / $\delta\nu_{k,2}$ = 0.6) for HS 0507+0434B. Based on the values of $\delta\nu_{k,1}$ and $\delta\nu_{k,2}$, we calculate some $m$ = $\pm$1 components for $l$ = 1 modes and some $m$ = $\pm$1 and $\pm$2 components for $l$ = 2 modes. With components of $l$ = 1 and 2, the independent mode of 999.7\,s, 6 further signals and 5 linear combinations can be fitted in Table 4. The short modes of 197.7\,s and 229.9\,s can also be fitted by components of $l$ = 1 and 2.

\begin{table*}
\begin{center}
\begin{tabular}{lllllllllllll}
\hline
$P_{\rm cal(s)}(l,k,m)$&$P_{\rm obs}$&$P_{obs}$-$P_{cal(s)}$&$P_{\rm cal(s)}(l,k,m)$&$P_{\rm obs}$&$P_{obs}$-$P_{cal(s)}$&$P_{\rm cal(s)}(l,k,m)$&$P_{\rm obs}$&$P_{obs}$-$P_{cal(s)}$\\
(s)              &(s)      &(s)         &(s)            &(s)      &(s)         &(s)             &(s)      &(s)         \\
\hline
196.91(1,1,1)    &         &            &228.61(2,5,-1) &         &            &752.78(2,23,-1) &         &            \\
197.05(1,1,0)    &         &            &228.92(2,5,-2) &229.9    &$\,$ 0.98   &756.19(2,23,-2) &         &            \\
197.19(1,1,-1)   &197.7    &$\,$ 0.51   &273.59(2,6,0)  &         &            &776.80(2,24,2)  &773.8    &$\,$-3.00   \\
231.07(1,2,1)    &229.9    &$\,$-1.17   &297.84(2,7,2)  &         &            &780.42(2,24,1)  &         &            \\
231.26(1,2,0)    &         &            &298.37(2,7,1)  &         &            &784.08(2,24,0)  &         &            \\
231.45(1,2,-1)   &         &            &298.90(2,7,0)  &         &            &787.77(2,24,-1) &         &            \\
294.64(1,3,0)    &         &            &299.44(2,7,-1) &         &            &791.50(2,24,-2) &         &            \\
354.47(1,4,0)    &355.3    &$\,$ 0.83   &299.97(2,7,-2) &301.3    &$\,$ 1.33   &812.62(2,25,0)  &         &            \\
394.70(1,5,0)    &         &            &321.19(2,8,0)  &         &            &842.59(2,26,0)  &         &            \\
445.22(1,6,0)    &445.3    &$\,$ 0.08   &358.33(2,9,2)  &         &            &870.30(2,27,0)  &         &            \\
513.46(1,7,0)    &         &            &359.10(2,9,1)  &         &            &895.63(2,28,0)  &         &            \\
553.05(1,8,0)    &556.5    &$\,$ 3.45   &359.87(2,9,0)  &         &            &924.27(2,29,0)  &         &            \\
598.76(1,9,0)    &         &            &360.65(2,9,-1) &361.0    &$\,$ 0.35   &954.52(2,30,0)  &         &            \\
658.71(1,10,0)   &655.9    &$\,$-2.81   &361.42(2,9,-2) &         &            &971.44(2,31,2)  &972.2    &$\,$ 0.76   \\
699.42(1,11,0)   &697.6    &$\,$-1.82   &384.27(2,10,0) &         &            &977.12(2,31,1)  &         &            \\
747.15(1,12,0)   &748.6    &$\,$ 1.45   &409.00(2,11,0) &         &            &982.86(2,31,0)  &         &            \\
798.26(1,13,0)   &         &            &443.59(2,12,0) &         &            &988.67(2,31,-1) &         &            \\
842.09(1,14,0)   &         &            &471.39(2,13,0) &         &            &994.55(2,31,-2) &         &            \\
889.00(1,15,0)   &         &            &495.58(2,14,0) &         &            &1000.51(2,32,2) &999.7    &$\,$-0.81   \\
933.53(1,16,0)   &         &            &528.68(2,15,0) &         &            &1006.53(2,32,1) &         &            \\
980.26(1,17,0)   &         &            &560.96(2,16,0) &         &            &1012.63(2,32,0) &         &            \\
1030.09(1,18,0)  &         &            &583.76(2,17,0) &         &            &1018.80(2,32,-1)&         &            \\
1082.18(1,19,0)  &         &            &616.81(2,18,0) &         &            &1025.04(2,32,-2)&         &            \\
1129.21(1,20,0)  &         &            &638.78(2,19,2) &         &            &1045.03(2,33,0) &         &            \\
1181.67(1,21,0)  &         &            &641.23(2,19,1) &         &            &1075.10(2,34,0) &         &            \\
1228.70(1,22,0)  &         &            &643.70(2,19,0) &         &            &1107.10(2,35,0) &         &            \\
1275.10(1,23,0)  &         &            &646.19(2,19,-1)&645.4    &$\,$-0.79   &1137.39(2,36,0) &         &            \\
1326.68(1,24,0)  &         &            &648.69(2,19,-2)&         &            &1166.18(2,37,0) &         &            \\
1371.16(1,25,0)  &         &            &659.42(2,20,2) &658.3    &$\,$-1.12   &1197.92(2,38,0) &         &            \\
1418.27(1,26,0)  &         &            &662.03(2,20,1) &         &            &1228.96(2,39,0) &         &            \\
1476.63(1,27,0)  &         &            &664.66(2,20,0) &         &            &1258.74(2,40,0) &         &            \\
113.80(2,1,0)    &         &            &667.31(2,20,-1)&         &            &1291.25(2,41,0) &         &            \\
134.14(2,2,0)    &         &            &669.99(2,20,-2)&         &            &1321.25(2,42,0) &         &            \\
196.31(2,3,2)    &         &            &693.88(2,21,2) &         &            &1351.38(2,43,0) &         &            \\
196.54(2,3,1)    &         &            &696.78(2,21,1) &         &            &1362.71(2,44,2) &         &            \\
196.77(2,3,0)    &         &            &699.69(2,21,0) &         &            &1373.91(2,42,1) &         &            \\
197.00(2,3,-1)   &         &            &702.63(2,21,-1)&703.9    &$\,$ 1.33   &1385.29(2,44,0) &1382.7   &$\,$-2.59   \\
197.23(2,3,-2)   &197.7    &$\,$ 0.47   &705.60(2,21,-2)&         &            &1396.86(2,44,-1)&         &            \\
207.71(2,4,0)    &         &            &724.60(2,22,0) &         &            &1408.63(2,44,-2)&         &            \\
227.68(2,5,2)    &         &            &742.75(2,23,2) &737.5    &$\,$-5.25   &1417.84(2,45,0) &         &            \\
227.99(2,5,1)    &         &            &746.07(2,23,1) &         &            &1449.92(2,46,0) &         &            \\
228.30(2,5,0)    &         &            &749.41(2,23,0) &         &            &1479.48(2,47,0) &         &            \\
\hline
\end{tabular}
\end{center}
\caption{The detailed fitting results with the $1(s)_{ref}$ model in Table 2.}
\end{table*}

 \subsection{The effect of screened Coulomb potential}

\begin{figure}
\begin{center}
\includegraphics[width=9.0cm,angle=0]{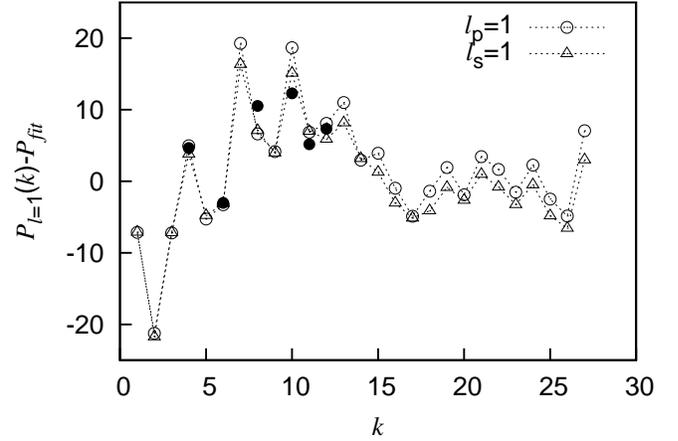}
\end{center}
\caption{The fitting diagram for $l$ = 1 modes. The two models are same as in Table 3. The open dots are calculated from the model evolved by adopting the pure Coulomb potential, while the triangles are calculated from the model evolved by adopting the screened Coulomb potential. The 27 ($k$ = 1-27) $l$ = 1 modes from the screened Coulomb potential model are fitted by $P_{fit}$ = 48.826*$k$ + 155.352. The figure is drawn by subtracting the fitting function from both the observed modes and the calculated modes. The observed modes are represented by filled dots.}
\end{figure}

\begin{figure}
\begin{center}
\includegraphics[width=9.0cm,angle=0]{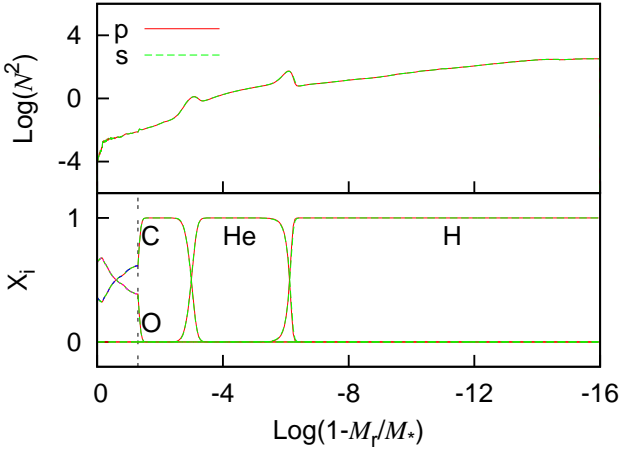}
\end{center}
\caption{The core composition profiles and the Brunt-V\"ais\"al\"a frequency for the models the same as in Table 3.}
\end{figure}

\begin{figure}
\begin{center}
\includegraphics[width=9.0cm,angle=0]{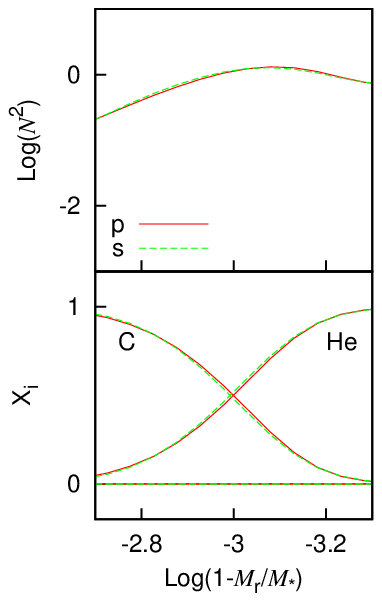}
\end{center}
\caption{The same as Fig. 3 but it enlarges the C/He transition zone.}
\end{figure}

\begin{figure}
\begin{center}
\includegraphics[width=9.0cm,angle=0]{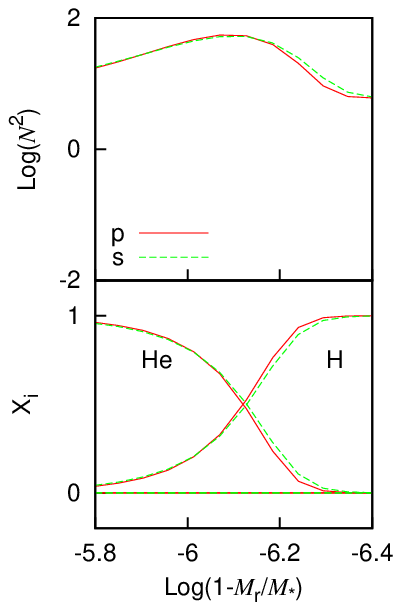}
\end{center}
\caption{The same as Fig. 3 but it enlarges the He/H transition zone.}
\end{figure}

\begin{figure}
\begin{center}
\includegraphics[width=9.0cm,angle=0]{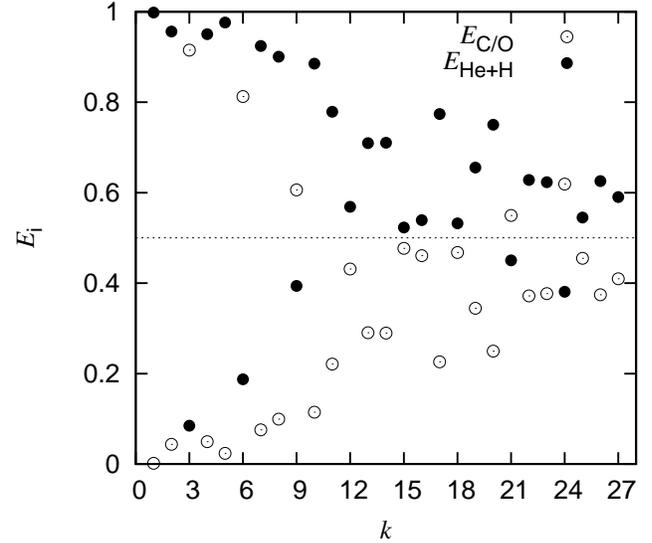}
\end{center}
\caption{The kinetic energy distribution diagram. The open dots represent the percentage of kinetic energy distributed in the C/O core, while the filled dots represent the percentage of kinetic energy distributed in the He and H envelope.}
\end{figure}

\begin{figure}
\begin{center}
\includegraphics[width=9.0cm,angle=0]{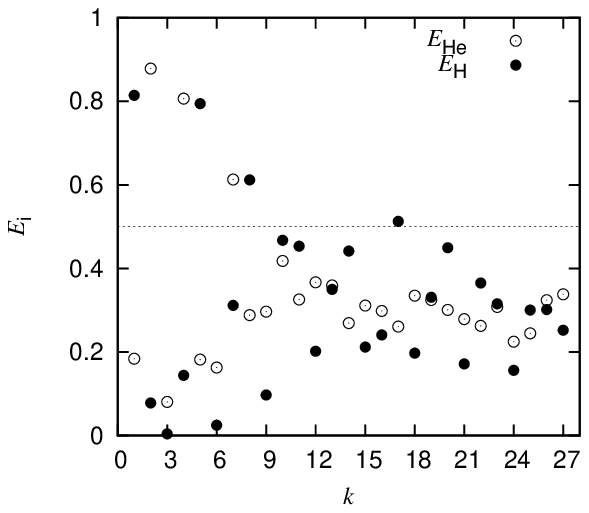}
\end{center}
\caption{The kinetic energy distribution diagram. The open dots represent the percentage of kinetic energy distributed in the He layer, while the filled dots represent the percentage of kinetic energy distributed in the H atmosphere.}
\end{figure}

In Fig. 2, we show the fitting diagram with two models the same as in Table 3. The abscissa is the radial overtone $k$. We calculate the mode with periods less than 1500\,s. The radial overtone is from 1 to 27. The open dots are calculated from the model evolved by adopting the pure Coulomb potential, while the triangles are calculated from the model evolved by adopting the screened Coulomb potential. The 27 triangles are fitted by $P_{fit}$ = 48.826*$k$ + 155.352. The figure is drawn by subtracting the fitting function from both the observed modes and the calculated modes. The observed modes are represented by filled dots. When fitting filled dots of $k$ = 10, the triangle is much better than the open dot. In addition, we can see that most of the high $k$ triangles are less than corresponding high $k$ open dots. Considering the screened Coulomb potential, the periods of most modes decrease. The periods of only few modes increase, such as the modes of $k$ = 5, 8, and 14.

In Fig. 3, we show the core composition profiles and the Brunt-V\"ais\"al\"a frequency for the models the same as in Table 3. The red lines represent the model evolved with pure Coulomb potential, while the green lines represent the model evolved with screened Coulomb potential. The differences between the red and green lines are very tiny in Fig. 3. On the left side of the vertical dashed line, we draw the core compositions of C (a blue line) and O (a pink line) of the white dwarf evolved by \texttt{MESA}. The C (O) profiles are overlapped in the core. In order to obtain a C/He interface in \texttt{WDEC}, we set the C abundance to 1.0, and then the O abundance to 0.0, at the surface of the core.

In Fig. 4, we enlarge Fig. 3 at the C/He transition zone. In Fig. 5, we enlarge Fig. 3 at the He/H transition zone. In Fig. 4, we can see that when adopting the screened Coulomb potential, the intersection of C/He abundances moves a little to the center of the star. Namely, the He and H envelope increase a little. In Fig. 5, we can see that when adopting the screened Coulomb potential, the intersection of He/H abundances moves a little to the surface of the star. Namely, the H atmosphere decreases a little.

In Fig. 6, we show a kinetic energy ($E_{i}$) distribution diagram. The kinetic energy distributions can be calculated by,
\begin{equation}
\begin{split}
 & \quad E_{C/O} \quad / \quad E_{He} \quad / \quad E_{H}\\
=& \quad {4\pi\int_{0}^{R_{C/He}}[(|\xi_{r}|^2+l(l+1)|\xi_{h}|^2)]\rho_{0}r^2dr} \\
/& \quad {4\pi\int_{R_{C/He}}^{R_{He/H}}[(|\xi_{r}|^2+l(l+1)|\xi_{h}|^2)]\rho_{0}r^2dr}\\
/& \quad {4\pi\int_{R_{He/H}}^{R}[(|\xi_{r}|^2+l(l+1)|\xi_{h}|^2)]\rho_{0}r^2dr}.
\end{split}
\end{equation}
\noindent In Eq.\,(4), $\rho_{0}$ is the local density and $R$ is the stellar radius. $R_{C/He}$ is the location of C/He interface and $R_{He/H}$ is the location of He/H interface. $E_{C/O}$, $E_{He}$, and $E_{H}$ is the value of kinetic energy distributed in the C/O core, the He layer, and the H atmosphere respectively. In Fig. 6, we can see that most modes have more than 50\% of their kinetic energy distributed in the He and H envelope, especially for high $k$ modes. According to the theory of Tassoul, Fontaine \& Winget (1990), $P_{k}$ $\infty$ $k$ ($\int$ $(N/r)$ dr)$^{-1}$, the integration is in the propagation region of modes.

Most modes have more than 50\% of their kinetic energy distributed in the He and H envelope, especially for high $k$ modes. When adopting the screened Coulomb potential, the He and H envelope increases as shown in Fig. 4. Therefore, the integration for most modes increase a little and the periods for most modes decrease a little. That is why most triangles are less than the corresponding open dots in Fig. 2.

In Fig. 7, we show the kinetic energy distribution diagram for the He layer and the H atmosphere. For the modes of $k$ = 1, 5, 8, 11, 14, 17, and 20, there are considerable amount of kinetic energy distributed in the H atmosphere. They are trapped or partly trapped in the H atmosphere. When adopting the screened Coulomb potential, the H atmosphere decreases as shown in Fig. 5. Therefore, the integration for those modes decrease a little and the periods should increase a little. In Fig. 2, we can see that the triangles of $k$ = 1, 5, 8, 11, and 14 are truly greater than the corresponding open dots. For the mode of $k$ = 17 and 20, although the triangles are not greater than the open dots in Fig. 2, they are not too much smaller. For the modes of $k$ = 2, 4, and 7, there are considerable amount of kinetic energy distributed in the He layer. They are trapped or partly trapped in the He layer. When adopting the screened Coulomb potential, the He layer increases a little and the periods should decrease a little. In Fig. 2, the triangles of $k$ = 2, 4, and 7 are truly less than the corresponding open dots, although they are low $k$ modes.

The laws can be used to study the mode trapping properties for the observed modes. In Table 2, the observed mode of 655.9\,s is fitted by 662.28\,s adopting the pure coulomb potential and fitted by 658.71\,s adopting the screened coulomb potential. The mode of 658.71\,s is much smaller than the mode of 662.28\,s. The observed mode of 655.9\,s may have considerable amount of kinetic energy distributed in the He and H envelope. The kinetic energy distributed in the H atmosphere does not dominate for the observed mode of 655.9\,s.

In Table 3, the $P_{cal(p)}$ modes used to fit the observed modes of 355.3\,s, 655.9\,s, and 748.6\,s are a little large. The corresponding $P_{cal(s)}$ modes decrease. The $P_{cal(p)}$ modes used to fit the observed modes of 445.3\,s and 556.5\,s are a little small. The corresponding $P_{cal(s)}$ modes increase. It is a preliminary research work, and the subsequent work will be carried out using the new version of \texttt{WDEC}.

 \subsection{Comparisons between the best fitting model and the previous work}

\begin{table*}
\begin{center}
\begin{tabular}{llllllllllllllll}
\hline
ID     &$T_{\rm eff}$  &log\,$g$       &$M_{*}$                &log($M_{\rm H}/M_{*}$)  &log($M_{\rm He}/M_{*}$)&$\sigma_{RMS}$\\
       &(K)            &               &($M_{\odot}$)          &                        &                       &(s)           \\
\hline
1      &11630$\pm$200  &8.17$\pm$0.05  &0.71                   &                        &                       &              \\
2      &11488$\pm$18   &8.057$\pm$0.008&                       &                        &                       &              \\
3      &12290$\pm$186  &8.24$\pm$0.05  &0.75$\pm$0.03          &                        &                       &              \\
4      &12257$\pm$135  &8.10$\pm$0.06  &0.660$\pm$0.023        &-4.43to-4.12            &-1.92                  &0.94          \\
5      &12460          &               &0.675                  &-8.5                    &-2.0                   &4.20          \\
6      &11450          &8.088          &0.640                  &-6.0                    &-3.0                   &1.94          \\
7(s)   &11790          &8.066          &0.625                  &-6.1                    &-3.0                   &2.08          \\
\hline
\end{tabular}
\end{center}
\caption{Table of best fitting models. The ID number 1, 2, and 3 is from spectral results of Fontaine et al. (2003), Koester et al. (2009), and Gianninas et al. (2011) respectively. The ID number 4, 5, 6, and 7 is from asteroseismological results of Romero et al. (2012), Fu et al. (2013), Chen \& Li (2014b), and the present paper respectively.}
\end{table*}

In this section, we discuss the previous study on HS 0507+0434B and the comparisons between the present study and the previous work. Based on the optical spectroscopy, Fontaine et al. (2003) obtained $T_{\rm eff}$ = 11630$\pm$200\,K and log$g$ = 8.17$\pm$0.05 for HS 0507+0434B. In the ESO Supernova Ia Progenitor Survey (SPY), high-resolution spectra of more than 1000 white dwarfs were obtained, including HS 0507+0434B. By fitting the high-resolution spectra, Koester et al. (2009) obtained $T_{\rm eff}$ = 11488$\pm$18\,K and log$g$ = 8.057$\pm$0.008 for HS 0507+0434B. Based on the high signal-to-noise ratio optical spectra, Gianninas, Bergeron \& Ruiz (2011) studied over 1100 DAV white dwarfs. They reported $T_{\rm eff}$ = 12290$\pm$186\,K, log$g$ = 8.24$\pm$0.05, and $M_{\rm *}$ = 0.75$\pm$0.03\,$M_{\odot}$ for HS 0507+0434B.

With the white dwarf evolution code of \texttt{LPCODE}, Romero et al. (2012) evolved grids of DAV star models and did asteroseismological study on 44 bright DAV stars. For HS 0507+0434B, they reported their best fitting model of log($M_{\rm He}/M_{\rm *}$) = -1.92, log($M_{\rm H}/M_{\rm *}$) = -4.43to-4.12, $T_{\rm eff}$ = 12257$\pm$135\,K, $M_{\rm *}$ = 0.660$\pm$0.023\,$M_{\odot}$, and log$g$ = 8.10$\pm$0.06. They fit 4 observed modes of 355.8\,s, 446.2\,s, 555.3\,s, and 743.4\,s. For the best fitting model, $\sigma_{RMS}$ is 0.94\,s. However, the mode of 355.8\,s, 446.2\,s, and 555.3\,3 was $m$ = -1, -1, and +1 component respectively identified by Fu et al. (2013). The mode of 743.3\,s was not identified by Fu et al. (2013).

With homogeneous mixture of C and O in the degenerate core, Fu et al. (2013) evolved grids of DAV star models and did asteroseismological model fittings on HS 0507+0434B. They obtained a best fitting model of log($M_{\rm He}/M_{\rm *}$) = -2.0, log($M_{\rm H}/M_{\rm *}$) = -8.5, $T_{\rm eff}$ = 12460\,K, and $M_{\rm *}$ = 0.675\,$M_{\odot}$. Fitting the 6 $m$ = 0 modes, they obtained the best fitting model with $\sigma_{RMS}$ = 4.20\,s.

By linear fittings to the C profile of white dwarf models evolved by \texttt{MESA}, Chen \& Li (2014b) produced grids of DAV star models by \texttt{WDEC} and did asteroseismological study on HS 0507+0434B. They adopted the diffusion equilibrium profiles of C/He and He/H interfaces. They obtained a best fitting model of log($M_{\rm He}/M_{\rm *}$) = -3.0, log($M_{\rm H}/M_{\rm *}$) = -6.0, $T_{\rm eff}$ = 11450\,K, $M_{\rm *}$ = 0.640\,$M_{\odot}$, and log$g$ = 8.088. Fitting the 6 $m$ = 0 modes, they obtained the best fitting model with $\sigma_{RMS}$ = 1.94\,s.

In this paper, we add the core compositions of white dwarf models evolved by \texttt{MESA} to \texttt{WDEC}. In order to obtain a C/He interface in \texttt{WDEC}, we set the C abundance to 1.0 at the surface of the core. The screened Coulomb potential effect is added into the element diffusion scheme. Grids of DAV star models are evolved. The theoretical modes are calculated and used to fit the 6 observed $m$ = 0 modes. We obtain a best fitting model of log($M_{\rm He}/M_{\rm *}$) = -3.0, log($M_{\rm H}/M_{\rm *}$) = -6.1, $T_{\rm eff}$ = 11790\,K, $M_{\rm *}$ = 0.625\,$M_{\odot}$, and log$g$ = 8.066. For the best fitting model, $\sigma_{RMS}$ is 2.08\,s.

We list the spectral and asteroseismological results in Table 5. For the same star (HS 0507+0434B), different research methods should obtain the same stellar parameters. The log$g$ from our best fitting model is basically consistent with that from spectral result of Koester et al. (2009). They are only 0.1\% different. Fitting the 6 $m$ = 0 modes, $\sigma_{RMS}$ = 2.08\,s for our best fitting model is obviously smaller than $\sigma_{RMS}$ = 4.20\,s in the work of ID 5. The best fitting model parameters are basically consistent with each other for the work of ID 6 and 7(s). This is due to the fact that both the work of ID 6 and 7(s) adopt the C profile of white dwarf models evolved by \texttt{MESA} and both the work of ID 6 and 7(s) evolve the grids of DAV star models by \texttt{WDEC}. In the work of ID 6, the diffusion equilibrium profiles of C/He and He/H interfaces are adopted. In this study, we treat the C/He and He/H interface profiles with element diffusion scheme adopting the pure Coulomb potential and screen Coulomb potential. In addition, this work, and others have shown that even small differences in chemical profiles can lead to large differences in the periods.

\section{Discussion and conclusions}

In this paper, we evolve grids of DAV star models by \texttt{WDEC}. The core compositions are from white dwarf models evolved by \texttt{MESA}. Instead of taking the diffusion equilibrium profiles, we take the element diffusion scheme of Thoul, Bahcal \& Loeb (1994). The evolved DAV star models are treated as taking the pure Coulomb potential into account. We added the screened Coulomb potential into the element diffusion scheme of Thoul, Bahcal \& Loeb (1994) and evolved the other grids of DAV star models. The theoretical modes are calculated and used to fit 6 reliable $m$ = 0 modes for DAV star HS 0507+0434B which were identified by Fu et al. (2013).

According to the value of $\sigma_{RMS}$, we obtain a best fitting model marked as $1(s)_{ref}$ in Table 2. The detailed fitting results are shown in Table 3 and Table 4. We obtained a $\sigma_{RMS}$ = 2.08\,s with the evolutionary DAV star models when fitting the 6 observed $m$ = 0 modes. The best fitting model parameters are log($M_{\rm He}/M_{\rm *}$) = -3.0, log($M_{\rm H}/M_{\rm *}$) = -6.1, $T_{\rm eff}$ = 11790\,K, $M_{\rm *}$ = 0.625\,$M_{\odot}$, and log$g$ = 8.066. Compared with the pure Coulomb potential scenario, the $\sigma_{RMS}$ is reduced by 34\% after using the screened Coulomb potential scenario. The screened Coulomb potential is very useful for the DAV star HS 0507+0434B.

Considering the screened Coulomb potential, the He and H envelope become a little thicker. The periods of most modes decrease a little, especially for the high $k$ modes. The H atmosphere becomes a little thinner. The periods of modes extremely trapped in the H atmosphere increase a little, such as the modes of $k$ = 1, 5, and 8.

For the spectral study, Koester et al. (2009) reported $T_{\rm eff}$ = 11488$\pm$18\,K and log$g$ = 8.057$\pm$0.008 for HS 0507+0434B. The gravitational acceleration of our best fitting model is only 0.1\% different from the result of Koester et al. (2009). The parameters of our best fitting model are basically consistent with that of asteroseismological work of Chen \& Li (2014b).

\section{Acknowledgements}

The work is supported by the National Natural Science Foundation of China (Grant No. 11803004 for the study of screened Coulomb potential, Grant No. 11563001 for the study of white dwarf pulsations, Grant No. 11333006 for the asteroseismological study on white dwarfs). We are very grateful to Y. Li, Q. S. Zhang, W. K. Zong, and J. Su for their kindly discussion and suggestions.

\label{lastpage}

\end{document}